# Positional uncertainty and quality assurance of digital elevation change detection (DECD)

Chang Li [a*], Qi Meng [a], Dong Wei [b], Wenzhong Shi [c] and Ming Hao [d]

a Key Laboratory for Geographical Process Analysing & Modelling, Hubei, and College of Urban and Environmental Science, Central China Normal University, Wuhan, China

b School of Remote Sensing and Information Engineering, Wuhan University, Wuhan, China

c Joint Research Laboratory on Spatial Information, The Hong Kong Polytechnic University and Wuhan University, Wuhan and Hong Kong, China

d School of Environment Science and Spatial Informatics, China University of Mining and Technology, Xuzhou, China

* Corresponding author: e-mail: lcshaka@126.com && lichang@ccnu.edu.cn

**Abstract:** Studies on rapid change detection of large area urgently need to be extended from 2D image to digital elevation model (DEM) due to the challenge of changes caused by disasters. This research investigates positional uncertainty of digital elevation change detection (DECD) caused by different degrees of DEM complexity and DEM misregistration. Unfortunately, using three-sigma rule (3σR) for DECD is disturbed by accuracy of parameter estimation, which is affected by the outliers (i.e., varied DEM) from DEM differencing samples. Hence, to reduce the aforementioned uncertainty of DECD, we propose a new strategy of quality assurance, adaptively censored three-sigma rule (AC3σR), in which with the samples censored, outliers of global DEM differencing samples outside the standard deviations of the mean calculated by moment estimation are iteratively removed. Compared with the 3σR and censored three-sigma rule (C3σR) that is similar to AC3σR but without iteration for both simulation and real-world data experiments, the proposed global AC3σR method always exhibits the highest accuracies of DECD in terms of both the overall accuracies 0.99967, 0.98740 and kappa coefficients 0.99598, 0.81803 respectively, and the strongest robustness with a large convergence interval [0, 0.30010] under the simulated maximum registration error and most complex terrain complexity conditions.

## 1. Introduction

Traditional 2D change detection is a major and classic technique used in remote sensing (Singh 1989, Lu et al. 2004, Hussain et al. 2013, Salah et al. 2020) and has been successfully applied in many fields, including agriculture (El-Kawy et al. 2011, Xing et al. 2018, Elagouz et al. 2020), disaster detection (Yamazaki and Matsuoka 2007, Novellino et al. 2019) and dynamic environmental change monitoring (Collado et al. 2002, Langat et al. 2019, Slingsby et al. 2020). Because it cannot fully consider the change in elevation information in digital terrain analysis (DTA), 2D change detection has difficulty accurately detecting changes in an area with large terrain undulations and in areas affected by earthquakes (Li et al. 2014, Moya et al. 2020, Wei and Yang 2020), debris flows (Miura 2019), landslides (Nichol and Wong 2005, Adriano et al. 2020) and other major disasters. Thus, to break down the barriers of the spectral variability, perspective distortion and lack of 3D elevation information observed in traditional 2D image analysis (Qin et al. 2016) and to meet the requirements of high-accuracy geoscience analysis and DTA, it is necessary to investigate digital elevation change detection (DECD). Importantly, DECD also provides additional data sources for analysis, such as height and depth or 2.5D information. Therefore, DECD has important theoretical and applied research value in geographical information science (GIScience), especially in DTA, which is beneficial not only for enriching the theory and method of DECD but also for studying geological disasters and their secondary disasters.

The recent development of DECD techniques includes projection-based differences (Crispell et al. 2012), object-oriented detection or classification (Nebiker et al. 2014, Huang et al. 2017, Peng and Zhang 2017, Han et al. 2020), and digital elevation model (DEM) differencing (Tian et al. 2010, Stal et al. 2013). An existing challenge is the sensitivity of registration error and the complexity of terrain in the geographical environment. However, research on positional uncertainty and quality assurance of DECD has not been reported, even in the field of DTA in GIScience. Moreover, the following four major problems associated with uncertainty in change detection are key issues that must be resolved to improve DECD accuracy.

- Registration error. Registration can be classified as 2D registration and 3D registration. In particular, 2D registration is mainly implemented in image registration, which must be employed in traditional 2D change detection before image differencing or image ratio processing, and its accuracy has an important influence on the next step of change detection. The impact of inaccurate image registration on 2D change detection has been studied in detail (Gottesfeld Brown 1992, Zitova and Flusser 2003, Tondewad and Dale 2020, Jiang et al. 2020). Townshend et al. (1992) concluded that image registration of 0.20000 pixels or less is required to achieve a positional error of only 10%; thus, a high level of image registration must be achieved to obtain reliable change detection. Dai and Khorram (1998) indicated that false changes resulting from misregistration are spatially distributed mainly along the edges of remotely sensed images. This topic was further studied by Shi and Hao (2013), who used buffer analysis to show the relations between detection errors and image edges. However, current uncertainty studies of change detection have mainly focused on remotely sensed 2D images (Yu et al. 2008, Dawn et al. 2010, Ma et al. 2019). Similarly, 3D registration affects the positional (i.e., geometric) uncertainty of DECD, which is mainly utilized in point cloud registration and DEM registration. The first step of DECD is the DEM registration (i.e., coregistration) of two periods. DEM registration transforms the DEMs into a unified coordinate system to complete the overlay and has some related studies as follows: Akca (2010) proposed a method of 3D surface coregistration based on least square matching, and tested its effectiveness with terrain experiments; Nuth and Kääb (2011) proposed an DEM coregistration method for the study of glacier thickness change; and Cucchiaro et al. (2020) showed that coregistration affects the estimation of multi temporal geomorphic changes. After registration, DEM differencing can be carried out to detect the change (de Albuquerque et al. 2020). However, errors inevitably are produced in registration, which affects the accuracy of DECD (Yue et al. 2015). Thus, the study of 3D misregistration for DECD is worthy of comprehensive study.

- Accuracy assessment. It is an important part of DECD. The main accuracy indices of DECD include overall accuracy (OA), kappa coefficient (kappa), producer accuracy (PA), user accuracy (UA) (Lu et al. 2004). In addition, intersection over Union (IOU) (Jaccard 1912, Taha and Hanbury 2015) can be used to assess the positional accuracy of DECD. The above accuracy indices are widely used in accuracy evaluation of change detection (Foody 2010, Martinez-Izquierdo et al. 2019) and can be evaluated using error matrix (Chughtai et al. 2021). Therefore, the universality of the aforementioned accuracy indices is conducive to scientific and effective evaluation of DECD results.

- Quality assurance. The essence of quality assurance is to improve the quality of all factors involved in production. This study includes three major types of factors, i.e., accidental error, gross error (outliers) and systematic error in the DECD processing. The influence of outliers (i.e., truly changed areas) on parameter estimation for DECD has not been considered. For example, DEM differencing can be regarded as random errors (López 1997, Wechsler 2003, Wechsler and Kroll 2006, Mesa-Mingorance and Ariza-López 2020) that usually follow a normal distribution; however, actual changes do not follow this distribution in DECD due to outliers or mixtures of errors, e.g., samples in the regions of debris flows and landslides. Therefore, Ariza-López et al. (2019) proposed a quality assurance strategy for non-normal error. These outliers contaminate samples and lead to the uncertainty of parameter estimation in DECD. Hence, how to remove outliers to ensure these errors in DECD is worthy of exploration. Outlier removal is to robustly detect and remove varied DEM between two phases. ISO (ISO16269-4:2010(E)) proposes robust data analysis to solve outliers in univariate data, and provides many methods, including resistant estimation, robust estimation, order statistic, trimmed mean, median and quartile etc. Furthermore, computer vision proposes many outlier removal methods, including the random sample consensus (RANSAC) (Fischler and Bolles 1981), Bayesian sampling Consensus (BaySAC) (Botterill et al. 2009), and the least median of squares (LMedS) (Rousseeuw 1984) etc. However, RANSAC needs to set thresholds for specific problems. Incorrectly assuming that the degenerate configurations contains outliers may also lead to the failure of BaySAC sampling strategy (Botterill et al. 2009). When the proportion of outliers in the samples reaches or exceeds 50%, LMedS is no longer applicable. Therefore, it is particularly important to propose a new adaptive and robust algorithm for DECD.

- Adaptive threshold. The reliability of change detection is based on threshold selection in DEM differencing. This procedure usually requires many trials and considerable experience to manually set the threshold (Bruzzone and Serpico 1997, Bruzzone and Cossu 2003, Rosin and Ioannidis 2003, Solano-Correa et al. 2018). Adaptively determining the threshold is a challenging process, and a suitable method has not been reported to date. When dealing with this indispensable change detection step, methods for determining the threshold have included manual classifications (Melgani et al. 2002, Wu et al. 2017), statistical analysis (Rogerson 2002, Khanbani et al. 2020a, Zhao et al. 2020, Khanbani et al. 2020b) and adaptive threshold estimation (Bruzzone and Prieto 2000, Bazi et al. 2010, Solano-Correa et al. 2019). When applying statistics-based methods, the crucial step is automatically and credibly determining the discriminant threshold of change detection to reduce positional uncertainty and improve DECD quality assurance, which is worthy of in-depth study.

This paper seeks to effectively solve the aforementioned problems in DECD. The main innovation and contributions of this paper include the following.

- The uncertainty of DECD is quantized and analysed by a newly proposed method of 3D misregistration simulation. Levels of terrain complexity and 3D misregistration are simulated by introducing errors in DEM registration and DEM differencing for the first time.

- Based on the basic principle of moment estimation for DEM differencing samples, a new strategy of global quality assurance, namely, adaptively censored three-sigma rule (AC3σR), is initially proposed to reduce the uncertainty of DECD and to ensure the quality of DECD. Although the changing elevation (i.e., outliers) leads to DEM differencing samples contamination and uncertainties in accuracy, and the proposed AC3σR can improve the accuracy of DECD by automatically removing outliers. Moreover, through robust iterative moment estimation, the optimal parameter estimation results are obtained, the adaptive threshold convergence radius is large, and the DECD results are stable.

- A hypothesis of symmetric distribution for global DEM differencing samples is proposed and proven by the skewness and delta index based on the proposed AC3σR with unilaterally censored samples, so that the normal distribution condition for DECD is extended and relaxed to the approximate symmetric distribution. Compared with the local DEM differencing, that the global DEM differencing is more effective for real-world DECD is verified by us.

- The uncertainty caused by 3D misregistration is systematically analysed, and high-accuracy DECD is realized, thus providing a theoretical basis for improving the accuracy of DECD in practice.

## 2. Methodology

### 2.1 DEM misregistration simulation

In the simulation, the real elevation changes (i.e., outliers) are known. The Gaussian synthetic surface function (Zhou and Liu 2004, Shi and Tian 2006, Li et al. 2018a, Li et al. 2018b) that is used to generate simulated DEMs at different levels of terrain complexity is simplified as follows:

$$z = Axe^{-x^2-y^2} \qquad (1)$$

where $A$ is the parameter used to determine the topographic relief. The larger the $A$ value is, the steeper the simulated terrain.

Image registration is generally employed to correct geometric deformations in multitemporal images. However, in this paper, registration processing is achieved by the relations between the polynomial model and 3-dimensional ground control points (3DGCPs). The transformation model between the old DEM grid points and the new DEM grid points is as follows:

$$\begin{aligned} X_{Old} &= a_0 + a_1 X_{New} + a_2 Y_{New} + a_3 Z_{New} \\ Y_{Old} &= b_0 + b_1 X_{New} + b_2 Y_{New} + b_3 Z_{New} \\ Z_{Old} &= c_0 + c_1 X_{New} + c_2 Y_{New} + c_3 Z_{New} \end{aligned} \qquad (2)$$

where ($X_{Old}$, $Y_{Old}$, $Z_{Old}$) and ($X_{New}$, $Y_{New}$, $Z_{New}$) are the grid points corresponding to the old (i.e., old phase) DEM and the new (i.e., new phase) DEM. The coefficients $a_0$, $a_1$, $a_2$, $a_3$, $b_0$, $b_1$, $b_2$, $b_3$, $c_0$, $c_1$, $c_2$ and $c_3$ represent the polynomial coefficients of the transformation model, including rotation, translation and scaling relationships between the old DEM and new DEM. In this research, rotation and translation are considered because DEM registration errors may become more obvious for DEM translation and rotation, which can be simulated by moving along the *X, Y* and *Z* directions of the registered DEM corresponding to the original DEM. During registration, this paper discusses the simulation of registration errors by considering unidirectional translation and simultaneous translation. This process requires that the error in the overall accuracy should be within a certain root-mean-square error (RMSE) range.

When DEM misregistration is simulated, there are some errors, which are defined as follows. (1) Accidental errors (i.e., random errors) are measurement errors that cause inconsistent observation values when measurements are repeated. (2) Outliers are extreme values that reflect the presence of local changes in DTA. In DECD, the DEM change for parameter estimation is considered to be an outlier.

*2.2 3σR-based DECD*

The DEM differencing method, one of the most common methods used for change detection, is applied in this section. Analogously, DECD performs DEM differencing by subtracting one DEM from another in the same region at different phases. If $Z_{Old}(X, Y)$ and $Z_{New}(X, Y)$ are the two elevation values for different phases of the DEM, corresponding to $DEM_{Old}$ and $DEM_{New}$, respectively, then DEM differencing yields $F_{ZD}(X, Y)$ as follows:

$$F_{ZD}(X,Y) = Z_{New}(X,Y) - Z_{Old}(X,Y) \qquad (3)$$

The following hypotheses are defined. (1) $DEM_{Old}$ and $DEM_{New}$ are free of outliers; (2) Stability: The difference in the heights of $DEM_{Old}$ and $DEM_{New}$ (i.e., DEM differencing) without changes can be considered a normal distribution or generally symmetric distribution; (3) Change: As $DEM_{New}$ changes, the DEM difference can be considered a mixture (i.e., accidental errors and outliers) of distributions. The position where the DEM difference reaches a high value, i.e., varied elevation, can be considered an outlier. Essentially, almost all of the DEM differences exhibit an approximate symmetric distribution for the stable zones, and the others have higher values for the varied zones.

Further analysis can determine change from DEM differencing, before which a threshold value shall be determined. The determination of the threshold value is a research topic of interest, and many automation methods have been introduced to solve this problem. In this paper, the adaptive threshold method and the three-sigma rule (3σR) or 3$\sigma$-rule, a step-by-step iteration process used for the quality assurance of DECD, are employed to obtain the threshold. The 3σR method confirms the threshold as follows:

$$P\{|F_{ZD}(X,Y)-\hat{\mu}| \leq 3\hat{\sigma}\} = 1-\alpha \tag{4}$$

where $\hat{\mu}$ represents the mean elevation value of the DEM difference in formula (3), $\hat{\sigma}$ represents the standard deviation of the DEM difference, $F_{ZD}(X, Y)$. Under the condition of an approximate symmetric probability distribution for DEM differencing samples, $\alpha$ is the significance level. If $\alpha$ is represented by a normal distribution, it corresponds to 0.00250. The elevation changes that do not fall on the confidence binary can be regarded as an elevation change grid. If the DEM difference is aligned with the abovementioned conditions, then these grids are deemed to have changed. Then, determining parameters $\hat{\mu}$ and $\hat{\sigma}$ must be addressed. However, we cannot directly estimate $\hat{\mu}$ and $\hat{\sigma}$ from $F_{ZD}(X, Y)$ as calculated by formula (3) since these samples include outliers caused by changed elevation, e.g., earthquakes, debris flows and landslides. Therefore, the innovative DECD method with quality assurance is proposed as follows.

*2.3 C3σR-based and AC3σR-based DECD*

The DEM differencing method is used to detect 3D changes by 3σR iteration based on the censored sample and symmetric distribution, which removes all outliers of DEM differencing caused by changing elevation before estimating parameters $\hat{\mu}$ and $\hat{\sigma}$. The core concept of quality assurance is to ensure false positives and false negatives (in statistical hypothesis testing, these analogous concepts are known as type I and type II errors, where a positive result corresponds to rejecting the null hypothesis and a negative result corresponds to not rejecting the null hypothesis).

DEM errors can usually be simulated with a normal distribution (Carlisle 2005). If it is assumed that DEM follows normal distribution and there are no outliers in the data set, three sigma can be used for outlier detection (Daniel and Tennant 2001, Hoehle and Hoehle 2009). Assuming that the DEM difference of two phases follows an approximate symmetric distribution, to remove the influence of outliers (i.e., changed DEM), this paper combines the 3σR with the manual threshold method to determine the censored samples, and this method is referred to as censored three-sigma rule (C3σR). In this method, with samples unilaterally censored and bilaterally censored by the formula (5), outliers are removed by 3σR and one-time moment estimation without iteration. However, this method needs to manually specify the threshold, and its result is unstable.

To solve the above problems, this paper proposes performing AC3σR with an adaptive threshold to determine censored samples and solve the problem of the two types of errors based on binary searches. After the adaptive threshold is determined, the changes in DEM are identified as outliers with step-by-step iterations based on the symmetric distribution. The main flow chart of the AC3σR proposed for reliable DECD is as follows:

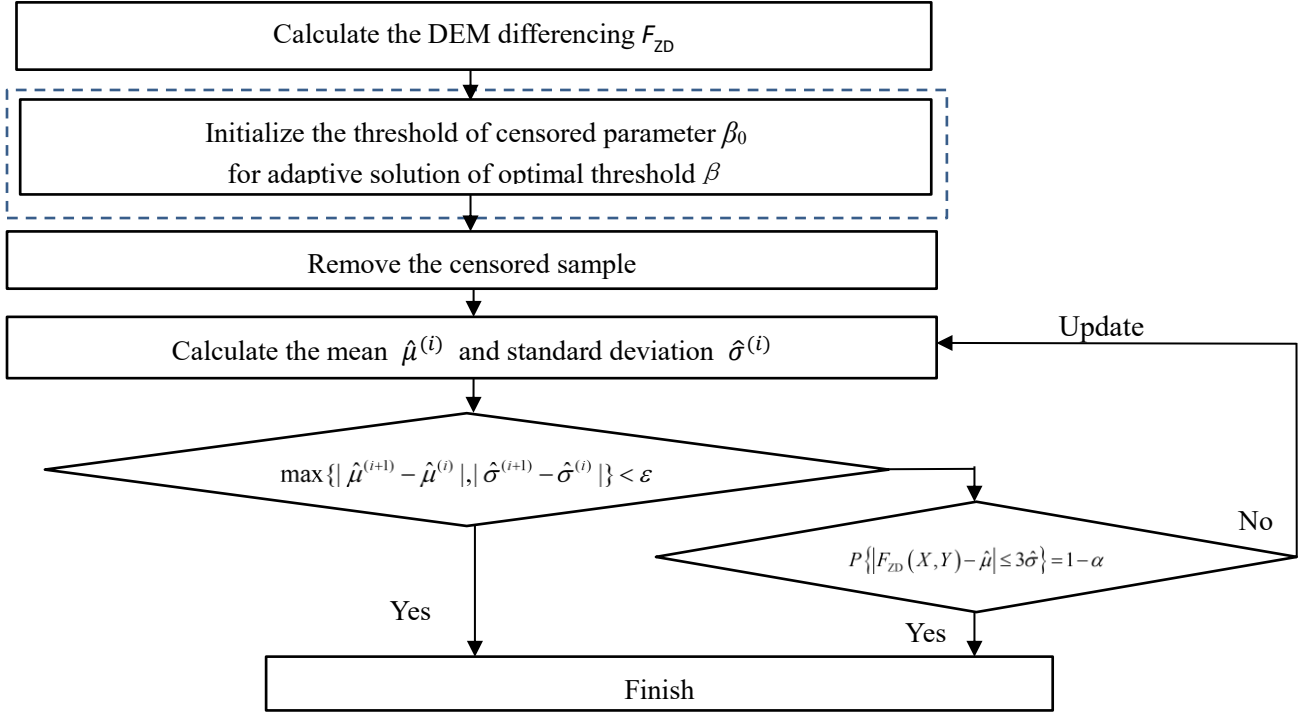

Figure 1. Flow chart for AC3σR (adaptively censored three-sigma rule).

In the flow chart, threshold $\beta$ is similar to a hyperparameter in machine learning and adaptive solution processing. Since the hyperparameter is given artificially and is insufficiently accurate, AC3σR can provide a stable radius of convergence, which can be proven by the experiment in Section 3.

The simple steps of AC3σR for DECD are as follows:

(1) The elements of a set are placed in ascending order

$$Error(i) = \begin{cases} \text{Sort}_{(i)}\{F_{ZD(i)}(X,Y)\}, \text{where } i=1,2,...,M \times N\} & \text{Bilateral processing} \\ \text{Sort}_{(i)}\{|F_{ZD(i)}(X,Y)|, \text{where } i=1,2,...,M \times N\} & \text{Unilateral processing} \end{cases} \quad (5)$$

Suppose that the overlapping area of the new-phase and old-phase DEM is $M$ (grid) × $N$ (grid) in size and the sample is $M \times N$ in quantity. For bilateral and unilateral processing, $Error(i)$ refers to the sort result of the $F_{ZD}$ or absolute value of $F_{ZD}$, respectively.

(2) The above sample data are censored by

$$\begin{cases} \text{Prob}\{Error(\beta/2) < F_{ZD(i)}(X,Y) < Error(MN-\beta/2)\} = 1-\beta/MN & \text{Bilateral censoring} \\ \text{Prob}\{|F_{ZD(i)}(X,Y)| < Error(MN-\beta/2)\} = 1-\beta/MN & \text{Unilateral censoring} \end{cases} \quad (6)$$

where $\beta$ is the outlier number, $\beta/MN$ is the outlier percentage and "Prob" represents the probability or frequency. This adjusted sample is pre-processed by bilateral censoring or unilateral censoring. Bilateral censoring means that both left-tailed samples and right-tailed samples of the $F_{ZD}$ are censored, and unilateral censoring means that right-tailed absolute value samples of the $F_{ZD}$ are censored. Please note that the unilateral censoring is equivalent to remove some large absolute value samples of $F_{ZD}$ (i.e., outliers or varied DEM), where the elevation increases and decreases obviously.

(3) Parameters are estimated based on moment estimation

According to the basic principle of moment estimation, it is assumed that the mean $\hat{\mu}$ and variance $\hat{\sigma}^2$ of DEM difference $F_{ZD}(X, Y)$ exist, and $F_{ZD(i)}(X, Y)$ is the $i$th grid sample of the DEM differences. The moment estimators of $\hat{\mu}_1$ and $\hat{\sigma}^2$ are as follows:

$$\begin{cases} \hat{\mu} = \hat{\mu}_1 = E(F_{ZD}(X,Y)) \\ \hat{\sigma}^2 = \hat{\mu}_2 - \hat{\mu}_1^2 = \frac{1}{n}\sum_{i=1}^{n}(F_{ZD(i)}(X,Y))^2 - [E(F_{ZD}(X,Y))]^2 = \frac{1}{n}\sum_{i=1}^{n}(F_{ZD(i)}(X,Y) - \hat{\mu})^2 \end{cases} \quad (7)$$

where $\hat{\mu}_1$ is the first sample moment of $F_{ZD}(X, Y)$, $\hat{\mu}_2$ is the second sample moment of $F_{ZD}(X, Y)$, and "$E$" is the expectation operator.

(4) The DEM change is detected

Whether DEM differencing follows the approximate symmetric probability distribution can be tested as follows.

1) The mean and median of the DEM difference are consistent, so the following formula is satisfied:

$$\Delta = \text{Mean}\{F_{ZD}(X,Y)\} - \text{Median}\{F_{ZD}(X,Y)\} = 0 \quad (8)$$

where $\Delta$ means subtracting the median of DEM differences (i.e., Mean{$F_{ZD}(X, Y)$})

from the mean of DEM differences (i.e., Median$\{F_{ZD}(X, Y)\}$).

2) The skewness of a symmetric distribution is equal to zero (Groeneveld and Meeden 1984) and is the third standardized moment $\hat{\mu}_3$, which is defined as follows:

$$\hat{\mu}_3 = E\left\{\left[(F_{ZD}(X,Y)-\hat{\mu})/\hat{\sigma}\right]^3\right\} = E\left[(F_{ZD}(X,Y)-\hat{\mu})^3\right] \Big/ \left(E\left[(F_{ZD}(X,Y)-\hat{\mu})^2\right]\right)^{3/2} \quad (9)$$

where $\hat{\mu}$ is the mean of $F_{ZD}(X, Y)$ and $\hat{\sigma}$ is the standard deviation of $F_{ZD}(X, Y)$.

(5) The iteration termination is determined

In the two nearest iterations, the mean $\hat{\mu}$ and standard deviation $\hat{\sigma}$ change very little, thus satisfying formula (10), where $\varepsilon$ is an arbitrarily small positive integer; then, the gradual iteration process converges. Otherwise, the existence of outliers can be judged by formula (4). If formula (4) is not met, then the inlier is updated, and the $\hat{\mu}$ and $\hat{\sigma}$ of the updated samples are recalculated. Then, whether formula (4) is met is determined again, and the iteration loop ends. In formula (10), the superscript $i$ represents the $i$th iteration.

$$\max\left\{|\hat{\mu}^{(i+1)} - \hat{\mu}^{(i)}|, |\hat{\sigma}^{(i+1)} - \hat{\sigma}^{(i)}|\right\} < \varepsilon \quad (10)$$

(6) The inliers of DEM difference $F_{ZD}$ are updated

The inliers and outliers of the DEM difference need to be updated to ensure the reliability of the inlier data before the censored samples are prepared for removal. If the samples satisfying formula (4) are moved from the inliers to the outliers, the $\hat{\mu}$ and $\hat{\sigma}$ after updating the inliers and outliers are recalculated, and formula (10) is reused for judgement.

## 2.4 Globally $F_{ZD}$-based DECD

At small scales (i.e., local $F_{ZD}$), some local DEM differencing samples follow approximate symmetric distribution, while the others may not follow symmetric distribution. For large scales or large samples, the local varied $F_{ZD}$ could be "averaged" by global $F_{ZD}$, so the global $F_{ZD}$ is closer to the approximate symmetric distribution than the local $F_{ZD}$. Therefore, DECD based on global $F_{ZD}$ is proposed in this paper, which can be proven and verified by both the simulated and real-world experiment in Section 3.

Actually, as long as an appropriate censored samples proportion for global $F_{ZD}$ is given, some abnormal DEM difference values (i.e., outliers or varied DEM) are removed by C3σR and AC3σR; then, that global $F_{ZD}$ approximately follows the symmetric distribution can be verified by the $\Delta$, $\hat{\mu}_3$, $\hat{\mu}_{(Global)}$ and $\hat{\sigma}_{(Global)}$ of global $F_{ZD}$.

### *2.5 Uncertainty evaluation of DECD*

In this paper, the traditional performance evaluation indices of the OA, kappa, PA, UA, and IOU are used. The IOU is defined as the area proportion of the intersection of the correctly detected DEM change boundary and the real DEM change boundary to the union set, so it can be used to evaluate the position accuracy. The confusion matrix and indices are as follows:

Table 1. The confusion matrix of DECD.

|  |  | Ground Truth | | |
|---|---|---|---|---|
|  |  | True | False | Total |
| Prediction | Positive | TP | FP | PP=TP+FP |
|  | Negative | FN | TN | PN=FN+TN |
|  | Total | GT=TP+FN | GF=FP+TN | TT=TP+FP+FN+TN |

Table 2. The performance indices of DECD.

| | OA | Kappa | PA | UA | IOU |
|---|---|---|---|---|---|
| Formula | $\dfrac{TP+TN}{TT}$ | $\dfrac{OA-(GT\times PP+GF\times PN)/TT^2}{1-(GT\times PP+GF\times PN)/TT^2}$ | $\dfrac{TP}{GT}$ | $\dfrac{TP}{PP}$ | $\dfrac{TP}{TP+FP+FN}$ |

Type I error corresponds to 1-UA, and type II error corresponds to 1-PA.

## 3. Experimental results and discussion

The overall experimental design is as follows: 1) a simulation data experiment is performed, data is introduced, and the performances of different algorithms, i.e., 3σR, C3σR and AC3σR are compared; 2) a real-world data experiment that is similar to the simulation data experiment, except for the data, is also performed; and 3) a discussion is provided, where the final results and mechanism of DECD are analysed, as detailed in the following sections.

### *3.1 Quality assurance of DECD for simulated data*

According to formula (1), different complex terrains are simulated. Figure 2 shows the change in steepness related to the parameters. According to the Gaussian synthetic surface function, *x*, *y* and *z* have no specific physical sizes in formula (1), so there is no specific physical size since these variables depend on the size of the display device and do not strictly correspond to an actual pixel or a physical unit.

Three simulated DEMs with different levels of complexity are created for the following experiment. The steepness of the simulated datasets is described in Table 3, and the ground truth of the change is shown in Figure 3, where different topographic complexity represents different geomorphic types: G1 simulates and represents a plateau, G2 represents a hill, and G3 represents a plain.

The change detection accuracy caused by 3D misregistration and terrain complexity is explored. To conduct this examination, the same elevation changes are added to the three simulated DEMs with different levels of complexity. Misregistration simulation adds errors between 0 and 1 to the transformation coefficients $a_0$, $a_1$, $a_2$, $a_3$, $b_0$, $b_1$, $b_2$, $b_3$, $c_0$, $c_1$, $c_2$ and $c_3$ in formula (2), which leads to translation, scaling and rotation errors. Then, 3D misregistration in the $x$, $y$ and $z$ directions occurs in each pair of simulated datasets. All datasets have the same columns and rows.

Data sets G1, G2 and G3 were used in the simulation experiment of DECD quality assurance. Figure 4 compares the DECD accuracy of 3σR, C3σR and AC3σR with different levels of terrain complexity. AC3σR determines its convergence threshold and radius by robust iterative estimation, and the iteration result is the result of the optimal convergence threshold experimentally tested.

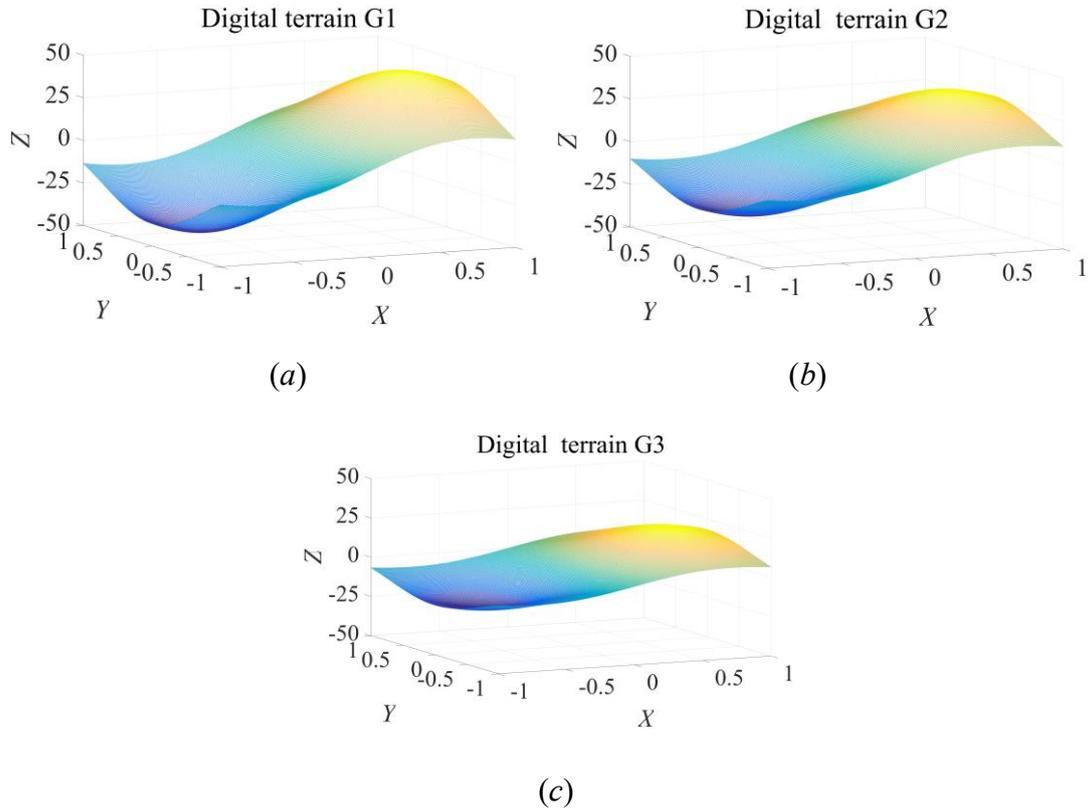

Figure 2. Simulated dataset. (*a*) *A*=100, (*b*) *A*=75, and (*c*) *A*=50.

Table 3. Simulated function parameter for the different terrains.

| Surface | G1 | G2 | G3 |
| --- | --- | --- | --- |
| A | 100 | 75 | 50 |
| Geomorphic types | Plateau | Hill | Plain |

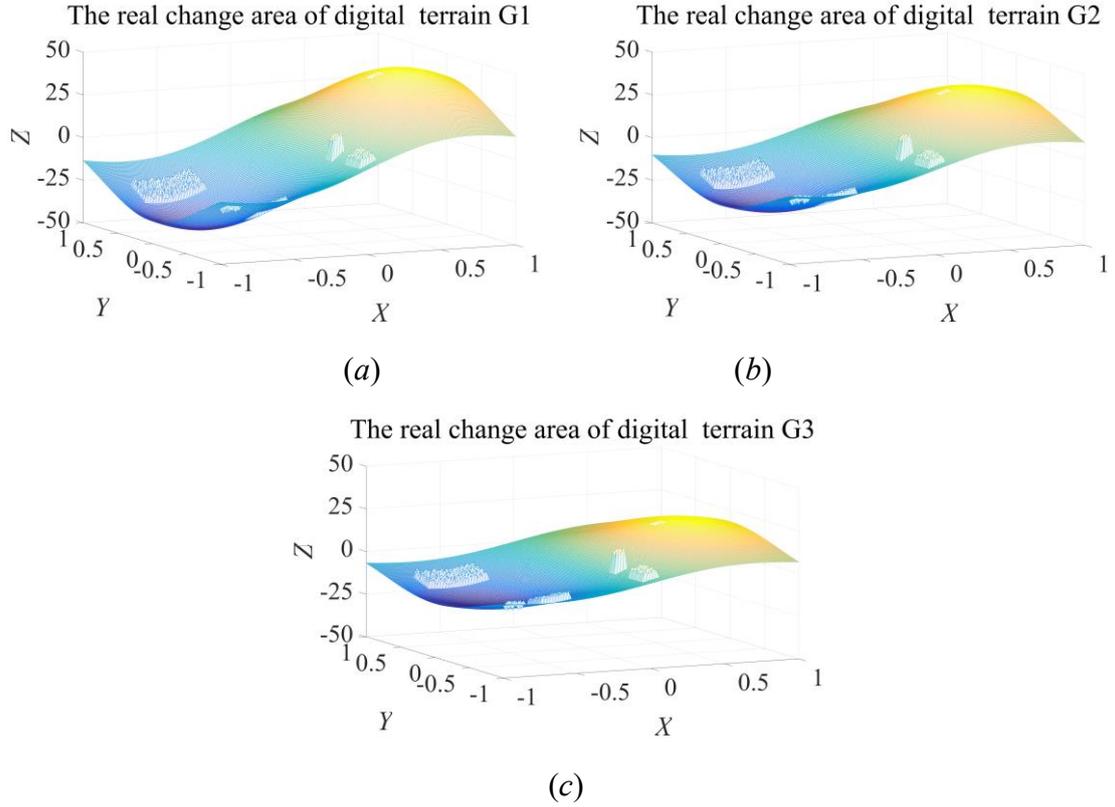

(a)   (b)

(c)

Figure 3. Ground truth of the simulated elevation change area corresponding to different terrain parameters, *A*. (a) *A*=100, (b) *A*=75, and (c) *A*=50, with changing elevation added by user-generated error.

First, to verify whether the DEM difference follows a symmetric distribution, this study calculated the Δ value in formula (8) and the $\hat{\mu}_3$ in formula (9) for terrains G1, G2 and G3, with the calculation results shown in Table 4, Table 5 and Table 6, respectively. The results show that Δ and $\hat{\mu}_3$ are close to zero, so the DEM difference follows an approximate symmetric distribution.

Table 4. Verification results of the approximate symmetric distribution of terrain G1.

| RE | 0.10000 | 0.30000 | 0.50000 | 0.70000 | 0.90000 |
|---|---|---|---|---|---|
| $\hat{\mu}$ | 3.97499 | 4.07972 | 4.03133 | 4.22312 | 4.59258 |
| Median | 4.06206 | 4.17754 | 4.09119 | 4.28136 | 4.57022 |
| Δ | 0.08708 | 0.09782 | 0.05987 | 0.05824 | -0.02236 |
| $\hat{\mu}_3$ | -0.08777 | -0.02514 | 0.00339 | -0.19639 | -0.54038 |

Table 5. Verification results of the approximate symmetric distribution of terrain G2.

| RE | 0.10000 | 0.30000 | 0.50000 | 0.70000 | 0.90000 |
|---|---|---|---|---|---|
| $\hat{\mu}$ | 3.98263 | 4.10258 | 4.04335 | 4.10348 | 4.53478 |
| Median | 4.06531 | 4.19224 | 4.18123 | 4.28105 | 4.58803 |
| Δ | 0.08267 | 0.08966 | 0.13787 | 0.17757 | 0.05325 |
| $\hat{\mu}_3$ | -0.08187 | -0.00051 | 0.02276 | -0.13162 | -0.32525 |

Table 6. Verification results of the approximate symmetric distribution of terrain G3.

| RE | 0.10000 | 0.30000 | 0.50000 | 0.70000 | 0.90000 |
|---|---|---|---|---|---|
| $\hat{\mu}$ | 3.99183 | 4.12550 | 4.04953 | 4.32004 | 4.58079 |
| Median | 4.07055 | 4.20898 | 4.26774 | 4.45708 | 4.66007 |
| Δ | 0.07872 | 0.08348 | 0.21821 | 0.13704 | 0.07927 |
| $\hat{\mu}_3$ | -0.07732 | 0.02392 | 0.03903 | -0.07051 | -0.11803 |

Then, this paper compares the effects of 3σR, C3σR and AC3σR with different terrain complexities and registration errors by both unilateral censoring and bilateral censoring. Because bilateral censoring for C3σR and AC3σR obtained better results than unilateral censoring in simulated experiments, only their comparison results are shown in Table 7 and Table 8.

3σR is less accurate than C3σR and AC3σR. When the registration errors (REs) increased, the OAs of C3σR and AC3σR were both higher than 0.97500, and their kappa values were both higher than 0.70000. However, when the registration error of the 3σR method was 0.90000 for terrain G1, the OA was only 0.96921, and the kappa value was only 0.45887. In addition, with the increase in terrain complexity, the decline in the OA and kappa value of 3σR was accelerated, and DECD did not show a significant advantage.

Within the allowed range of REs, the OA and kappa value of AC3σR were higher than those of C3σR overall. With the increase in REs, C3σR was slightly more accurate than AC3σR. The specific reasons for this will be analysed later.

To further explore C3σR and AC3σR in different terrain and registration error scenarios and assess the superiority of the DECD method, the value of hyperparameter $β$ was selected from 0 to 1 in this paper, and the influence of the adaptive threshold in AC3σR on the quality assurance of DECD was experimentally studied. Table 7 and Table 8 show comparisons of the highest OA and kappa values, respectively, achieved by C3σR and AC3σR with different terrain complexities. Figure 5, Figure 6 and Figure 7 correspond to the influence of the adaptive threshold on the detection accuracy of 3D change based on different registration errors when C3σR and AC3σR are applied to terrains G1, G2 and G3, respectively. The following results were obtained.

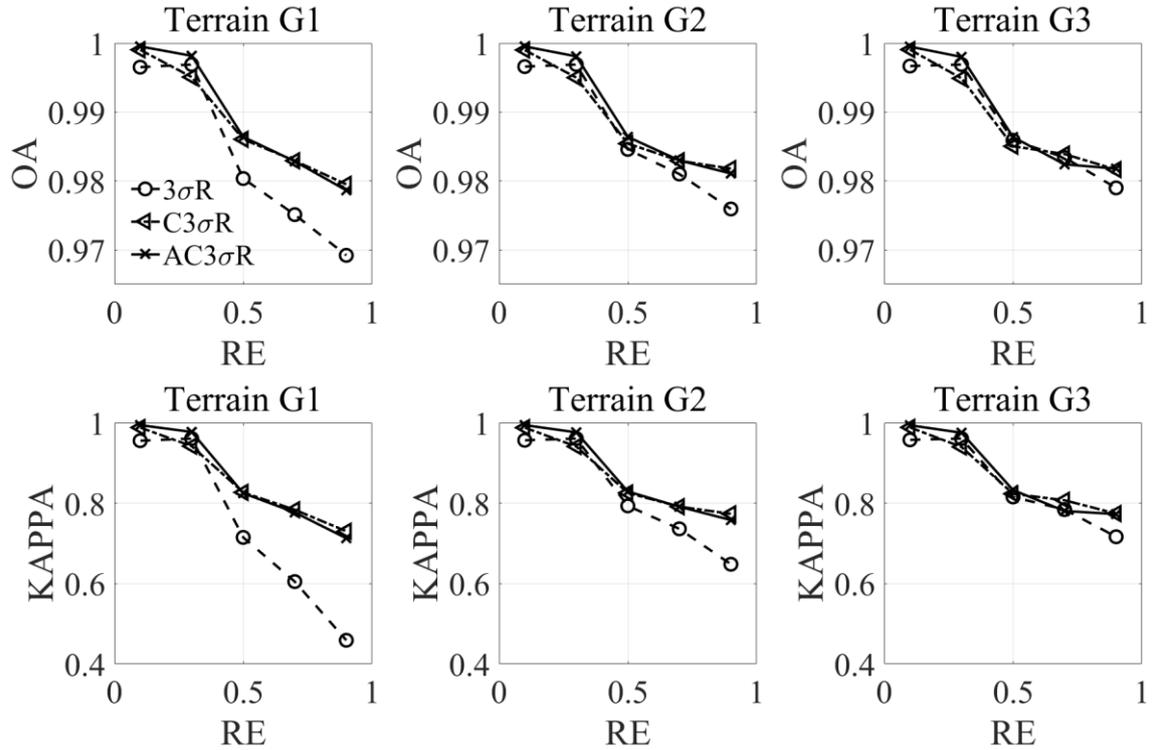

Figure 4. The accuracy of DECD with different REs (registration errors) for 3σR (three-sigma rule), C3σR (censored three-sigma rule) and AC3σR (adaptively censored three-sigma rule) under different terrain complexity conditions.

AC3σR is always more accurate than C3σR. The OA and kappa values of AC3σR

reach 0.99967 and 0.99598, respectively, while those of C3σR reach 0.99864 and 0.98355, respectively.

AC3σR is more stable than C3σR. The true change (i.e., outlier) ratio is 0.00399. When the censored samples proportion is selected near 0.00399, both AC3σR and C3σR can reach maximum accuracy to obtain the best DECD result. The censored sample proportion of the AC3σR method converges to the best accuracy within the convergence radius [0, 0.30010], while the censored samples proportion of C3σR converges to the best accuracy within the convergence radius [0, 0.05010]. When the proportion of censored samples exceeds the radius of convergence, the OA, kappa value and UA of AC3σR declines gently while the PA increase slowly, and the change trends show very good stability. However, the accuracy index of C3σR changes abruptly.

Table 7. Maximum OA of C3σR and AC3σR for different terrains.

|         | G1      |         | G2      |         | G3      |         |
|---------|---------|---------|---------|---------|---------|---------|
| RE      | C3σR    | AC3σR   | C3σR    | AC3σR   | C3σR    | AC3σR   |
| 0.10000 | 0.99864 | 0.99967 | 0.99854 | 0.99967 | 0.99854 | 0.99965 |
| 0.30000 | 0.99699 | 0.99756 | 0.99699 | 0.99756 | 0.99702 | 0.99745 |
| 0.50000 | 0.98608 | 0.98641 | 0.98513 | 0.98641 | 0.98595 | 0.98592 |
| 0.70000 | 0.98324 | 0.98334 | 0.98332 | 0.98375 | 0.98435 | 0.98454 |
| 0.90000 | 0.98177 | 0.98177 | 0.98166 | 0.98180 | 0.98166 | 0.98193 |

Table 8. Maximum kappa values of C3σR and AC3σR for different terrains.

|         | G1      |         | G2      |         | G3      |         |
|---------|---------|---------|---------|---------|---------|---------|
| RE      | C3σR    | AC3σR   | C3σR    | AC3σR   | C3σR    | AC3σR   |
| 0.10000 | 0.98355 | 0.99598 | 0.98225 | 0.99598 | 0.98225 | 0.99565 |
| 0.30000 | 0.96198 | 0.97006 | 0.96198 | 0.97006 | 0.96233 | 0.96877 |
| 0.50000 | 0.83105 | 0.83072 | 0.82465 | 0.83186 | 0.82280 | 0.82928 |
| 0.70000 | 0.79775 | 0.79638 | 0.79843 | 0.80578 | 0.81532 | 0.81868 |
| 0.90000 | 0.77287 | 0.77343 | 0.77320 | 0.77363 | 0.77459 | 0.77900 |

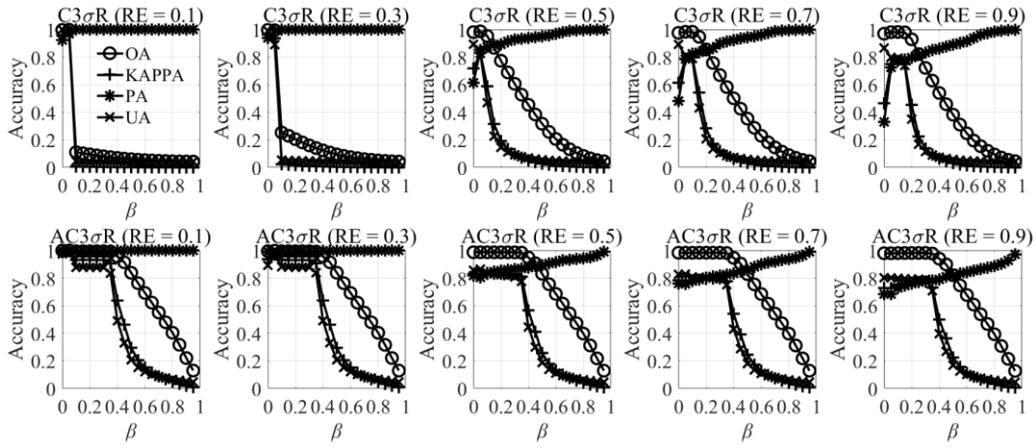

Figure 5. DECD accuracy of C3σR (censored three-sigma rule) and AC3σR (adaptively censored three-sigma rule) with different registration errors for terrain G1.

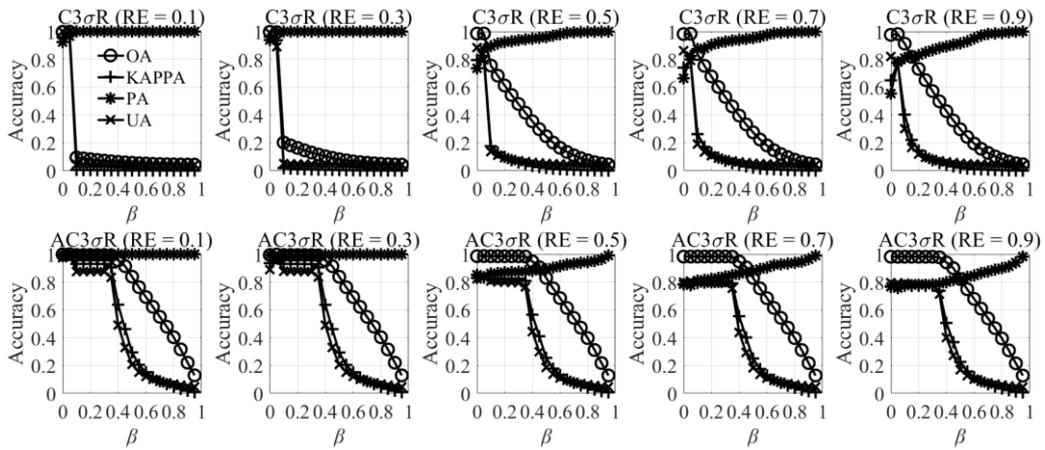

Figure 6. DECD accuracy of C3σR and AC3σR with different registration errors for terrain G2.

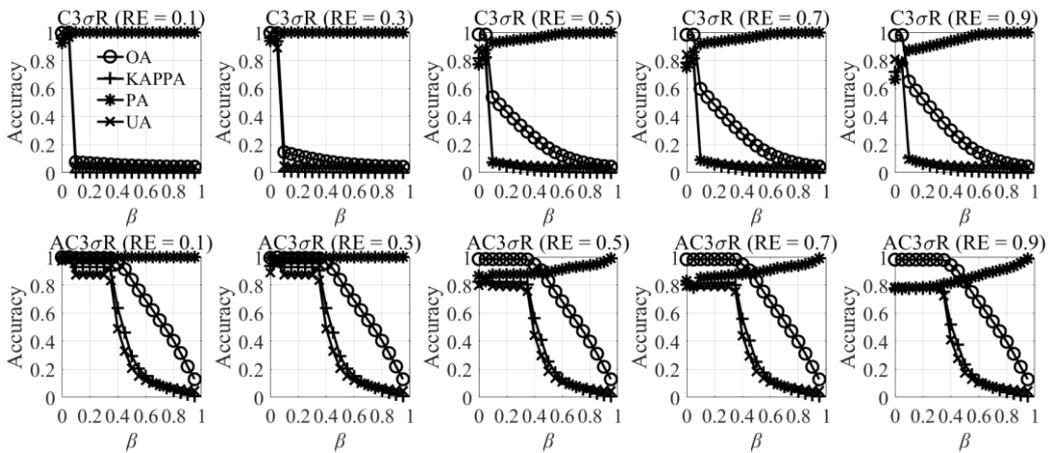

Figure 7. DECD accuracy of C3σR and AC3σR with different registration errors for terrain G3.

*3.2 Quality assurance of DECD for real-world data*

To study the universality of AC3σR, this paper carried out a DECD experiment based on data from Yingxiu town, Wenchuan County, Sichuan Province, China, in April 2010 and June 2011. In 2008, the year of the major earthquake in Sichuan, Yingxiu was at the epicentre, so 80% of it was destroyed. Two years later, on 14th August, 2010, secondary disasters, e.g., debris flows, occurred in Hongchun gully (N 31°04'01.1", E 103°29'32.7"), Fengxiangshu village, and Yingxiu town (Gan et al. 2012, Xu et al. 2012, Han et al. 2013, Li et al. 2015, Chen et al. 2016). The data were provided by the Earth Observation Centre of the Chinese Academy of Sciences and processed by DPGrid (digital photogrammetry grid) software. The resolution of the DEM is 5 m × 5 m. The overlaying digital orthophoto map (DOM) of the study area and the two controlled areas is shown in Figure 8. In sample area 1, we selected two local regions, as shown in Figure 8.(g) and Figure 8.(i). The profile diagrams corresponding to the profile lines are shown in Figure 8.(h) and Figure 8.(j). In the Figure 8.(h) and Figure 8.(j), the green one is the old phase DEM in 2010 and the blue one is the new phase DEM in 2011. The DEM differencing $F_{ZD}$ of the two phases calculated by formula (3) has positive and negative values. In the Figure 8.(h) and Figure 8.(j), the red region is the area of change detected by the proposed AC3σR using global samples. After the censored samples are removed from the 2.90000% global $F_{ZD}$ that means not the profile DEM differencing but the whole DEM differencing, the DECD result is calculated by the proposed AC3σR. Then, the $\hat{\mu}$ and $\hat{\sigma}$ of the global $F_{ZD}$ are used as parameters to detect local elevation changes, whose results are as follows:

(1) The number of samples about global $F_{ZD}$ that means the whole DEM differencing is 1140902, and the numbers of samples about local $F_{ZD}$ that means the profile DEM differencing are 36 in Figure 8.(g) and 13 in Figure 8.(i) respectively;

(2) The local $F_{ZD}$ are shown in Figure 8.(h) and Figure 8.(j), where $\Delta_L$ and $\hat{\mu}_{L3}$ represent the $\Delta$ calculated by formula (8) and skewness calculated by formula (9) of the local $F_{ZD}$ (i.e., the profile DEM differencing) are -2.60068, -0.71251 and 1.12836, 1.22597 respectively. Since the $\Delta$ and skewness $\hat{\mu}_3$ of the symmetric distribution are 0, and the skewness of the approximately symmetric distribution ranges from -0.50000 to 0.50000 (Joh and Malaiya 2014). So local $F_{ZD}$ of Figure 8.(h) and Figure 8.(j) cannot be regarded as the approximate symmetric distribution;

(3) Compared with the results of DECD based on local $F_{ZD}$, the better results of 3σR, C3σR and AC3σR based on global $F_{ZD}$ and unilateral censoring are shown in Table 9.

1) The DECD accuracies of local $F_{ZD}$ are very poor. As shown in Table 9, 3σR cannot detect any varied elevation of local area 1 at all. C3σR can only detect a part of changes in local area 1. Due to the $|\hat{\mu}_3|>0.5$ and $|\Delta|>1$ in both local area 1 and local area 2, the local $F_{ZD}$ does not follow the approximate symmetrical distribution.

2) With global $F_{ZD}$ samples censored or some outliers removed, both $\Delta$ and $\hat{\mu}_3$ of the global $F_{ZD}$ calculated by formula (8) and formula (9) are close to zero (i.e., -0.00533 and 0.00380 respectively), so the global DEM differencing samples follow the approximate symmetric distribution. AC3σR with global $F_{ZD}$ can obtain the best DECD accuracy.

It is worth noting that global $F_{ZD}$ is proposed for DECD, and even if the distribution of local $F_{ZD}$ does not follow the symmetric distribution, that of global distribution always follows the symmetric distribution, whose reasons are explained in discussion section.

Table 9. Comparison table of DECD accuracy of local area 1 and local area 2 based on local $F_{ZD}$ and global $F_{ZD}$.

|  |  | OA | kappa | PA | UA | IOU | $\hat{\mu}_3$ | Δ |
|---|---|---|---|---|---|---|---|---|
| Local area 1 (Local) | 3σR | 0.27778 | 0 | 0 | *nonexistent* | 0 | -0.71251 | -2.60068 |
| | C3σR | 0.41667 | 0.04545 | 0.26923 | 0.77778 | 0.25000 | -0.65740 | -1.24884 |
| | AC3σR | 0.66667 | 0.29870 | 0.65385 | 0.85000 | 0.58621 | -0.41843 | 0.29586 |
| Local area 2 (Local) | 3σR | 0.38462 | 0 | 0 | *nonexistent* | 0 | 1.22597 | 1.12836 |
| | C3σR | 0.38462 | 0 | 0 | *nonexistent* | 0 | 0.97693 | 0.90837 |
| | AC3σR | 0.38462 | 0 | 0 | *nonexistent* | 0 | 0.97693 | 0.90837 |
| Local area 1 (Global) | 3σR | 0.27778 | 0 | 0 | *nonexistent* | 0 | 0.21886 | 0.59964 |
| | C3σR | 0.61111 | 0.22222 | 0.57692 | 0.83333 | 0.51724 | 0.00380 | -0.00533 |
| | AC3σR | 0.86111 | 0.64286 | 0.92308 | 0.88889 | 0.82759 | -0.00219 | 0.00119 |
| Local area 2 (Global) | 3σR | 0.38462 | 0 | 0 | *nonexistent* | 0 | 0.21886 | 0.59964 |
| | C3σR | 0.76923 | 0.49351 | 0.87500 | 0.77778 | 0.70000 | 0.00380 | -0.00533 |
| | AC3σR | 0.76923 | 0.45070 | 1 | 0.72727 | 0.72727 | -0.00219 | 0.00119 |

The 3DGCPs of the DEM are obtained by the following photogrammetry technique. The corresponding points are obtained by image matching the two phases of the DOM, and then, the DOM is mapped or registered with the DEM so that the 3DGCPs can be obtained. The plane and elevation accuracy of 3DGCPs are about 0.75642m and 1.58247m, respectively. Positional accuracy is evaluated by 3D checkpoints from existing geographic information products, such as a DLG (digital line graphic), DOM, DEM and DSM (digital surface model), instead of field checkpoints.

Visual interpretation is completed by referring to two factors: (1) DEM change, whose change is detected according to the results of DEM differencing, and (2) land cover, whose change is detected according to the spectral and texture features (Carleer and Wolff 2006), such as those of debris flows and landslides.

Unilateral censoring for C3σR and AC3σR obtained better results than bilateral censoring in real-world experiments. The comparison of the experimental results of 3σR, C3σR and AC3σR with unilateral censoring is shown in Figure 10 and Figure 11, and the red areas are the areas of DEM change. The accuracy comparison table is shown in Table 10. The experimental results show that (1) AC3σR is more capable than C3σR and 3σR in the DECD of geological hazards such as landslides. (2) Compared with those of 3σR and C3σR, the results of AC3σR are more consistent with visual interpretation.

The DECD results are shown below from different perspectives. In addition, this paper selects two controlled areas in Figure 12 and Figure 13 to study the sensitivity of the DECD method in geological disaster detection. Table 10 shows that the DECD results of the AC3σR method for landslides, debris flows and other disasters are better than those of the other methods, i.e., 3σR and C3σR.

Table 10. Accuracy comparison table for C3σR and AC3σR.

|       | 3σR     | C3σR    | AC3σR   |
|-------|---------|---------|---------|
| OA    | 0.97670 | 0.98277 | 0.98740 |
| kappa | 0.66346 | 0.75077 | 0.81803 |
| PA    | 0.67609 | 0.75900 | 0.82511 |
| UA    | 0.67501 | 0.76042 | 0.82401 |
| IOU   | 0.60821 | 0.64396 | 0.73268 |

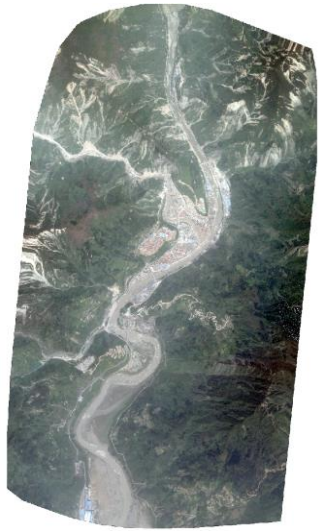 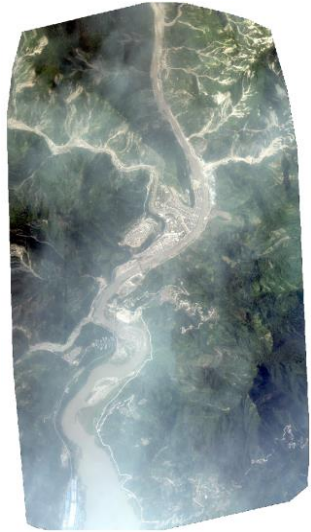

(a) Data for 2010   (b) Data for 2011

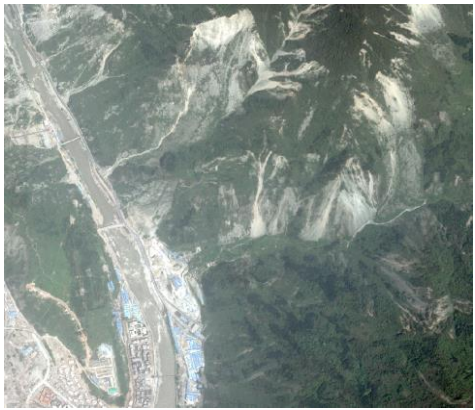 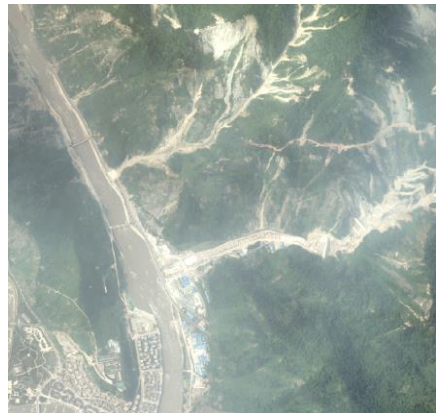

(c) Sample area 1 for 2010   (d) Sample area 1 for 2011

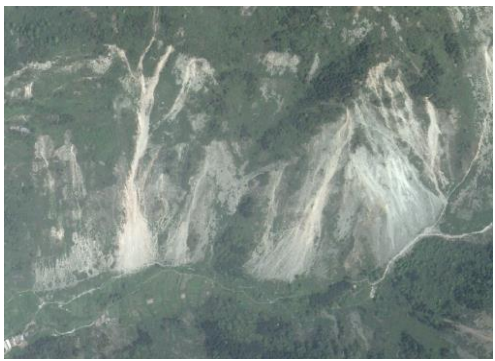 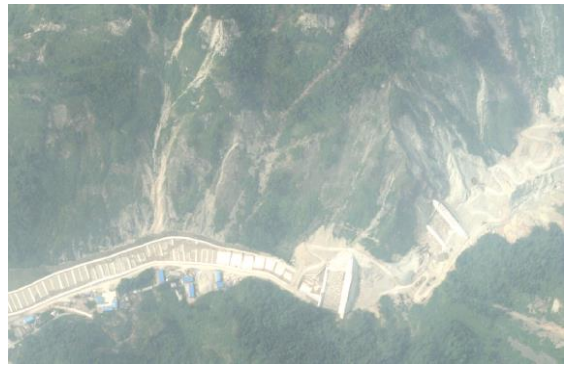

(e) Sample area 2 for 2010   (f) Sample area 2 for 2011

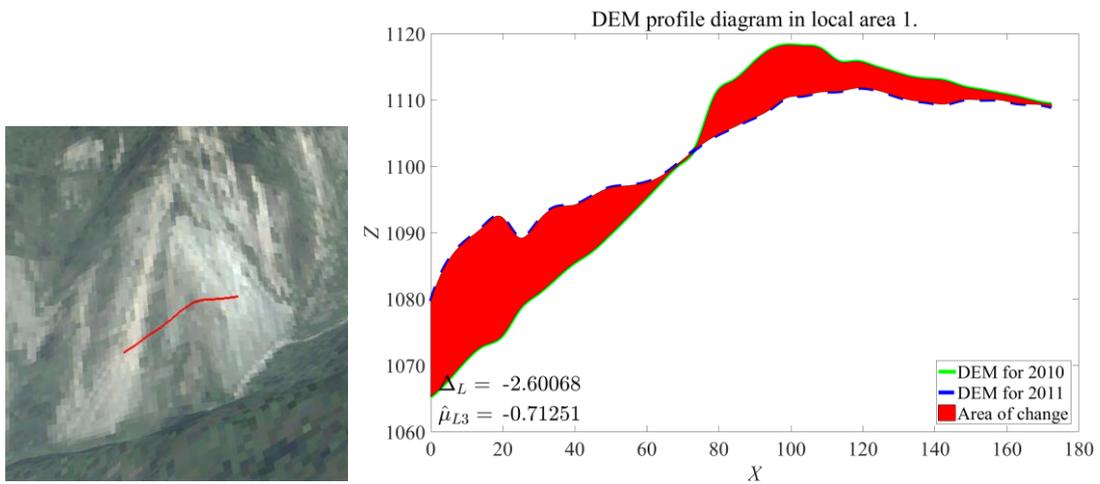

(g) DEM profile line in local area 1.  (h) DEM profile diagram in local area 1.

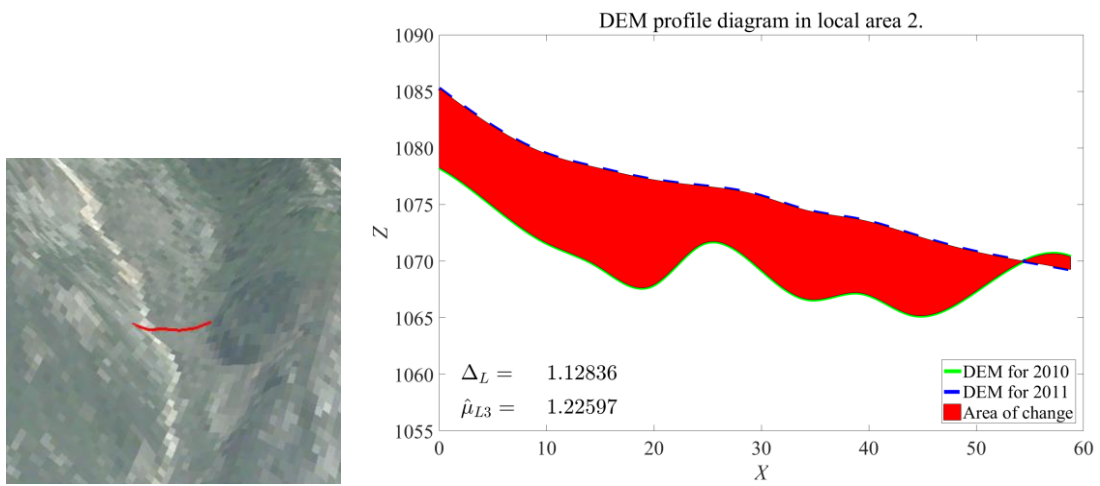

(i) DEM Profile line in local area 2.  (j) DEM profile diagram in local area 2.

Figure 8. Real data experimental area of Hongchun gully (N 31°04'01.1", E 103°29'32.7"), Fengxiangshu village, Yingxiu town, Wenchuan County, Sichuan Province, China.

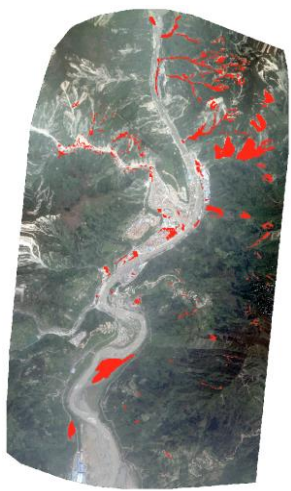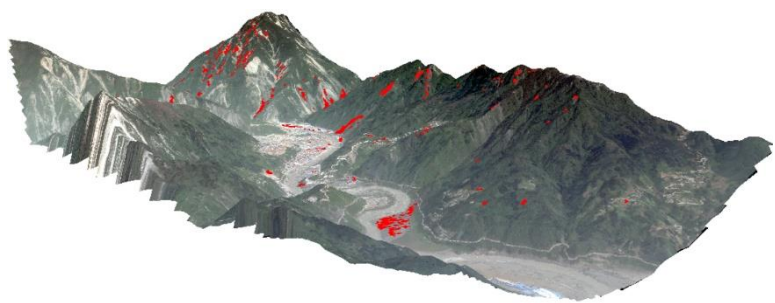

(a) 2D display  (b) 3D display

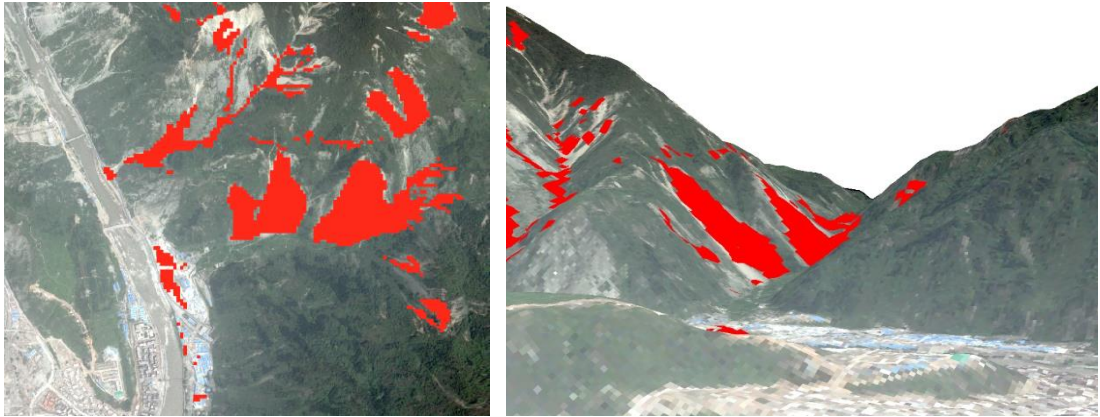

(c) 2D display of sample area 1   (d) 3D display of sample area 1

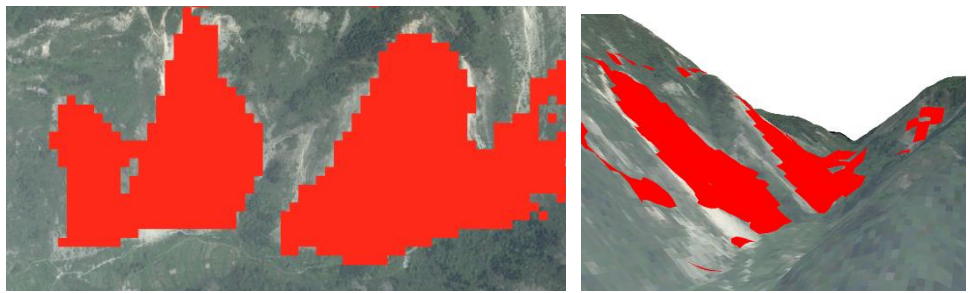

(e) 2D display of sample area 2   (f) 3D display of sample area 2

Figure 9. Visual interpretation of varied areas.

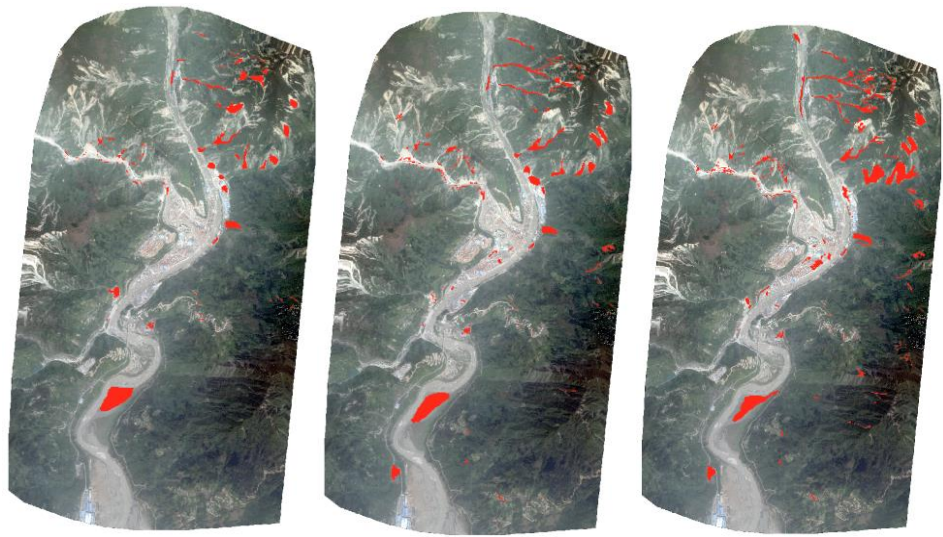

(a)　　　　　　　　(b)　　　　　　　　(c)

Figure 10. 2D display of real data experimental results obtained by (a) 3σR, (b) C3σR and (c) AC3σR.

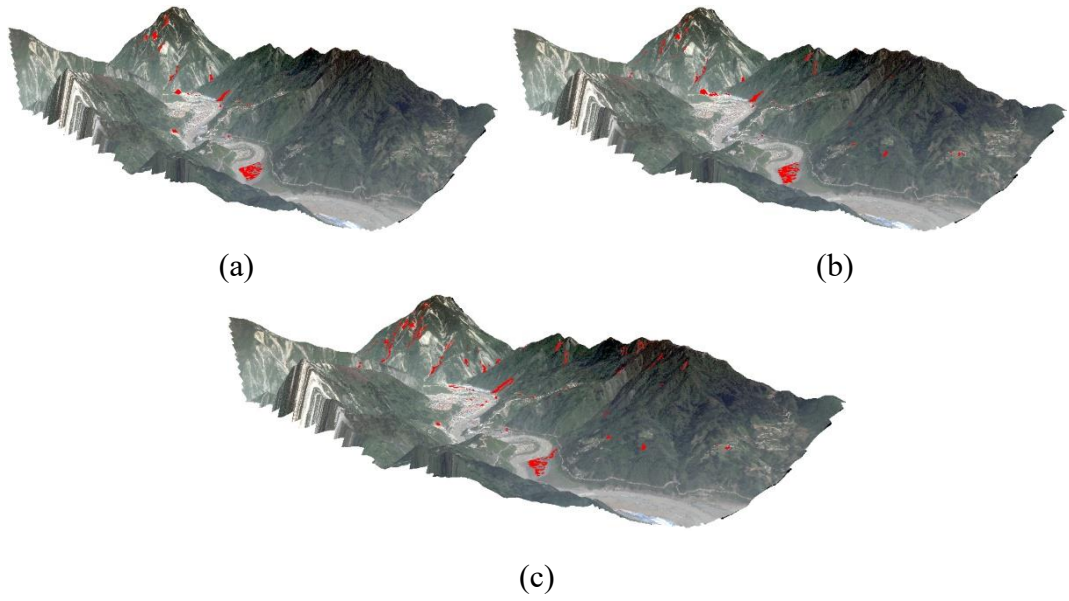

Figure 11. 3D display of real data experimental results obtained by (a) 3σR, (b) C3σR and (c) AC3σR.

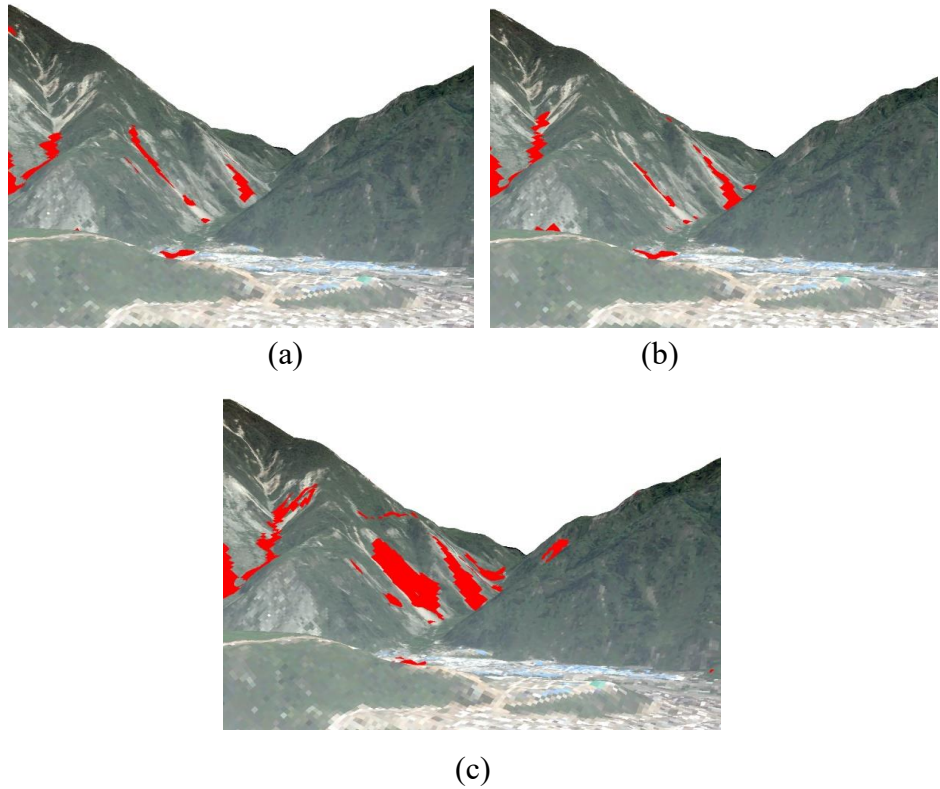

Figure 12. DECD results of sample area 1 obtained by (a) 3σR, (b) C3σR and (c) AC3σR.

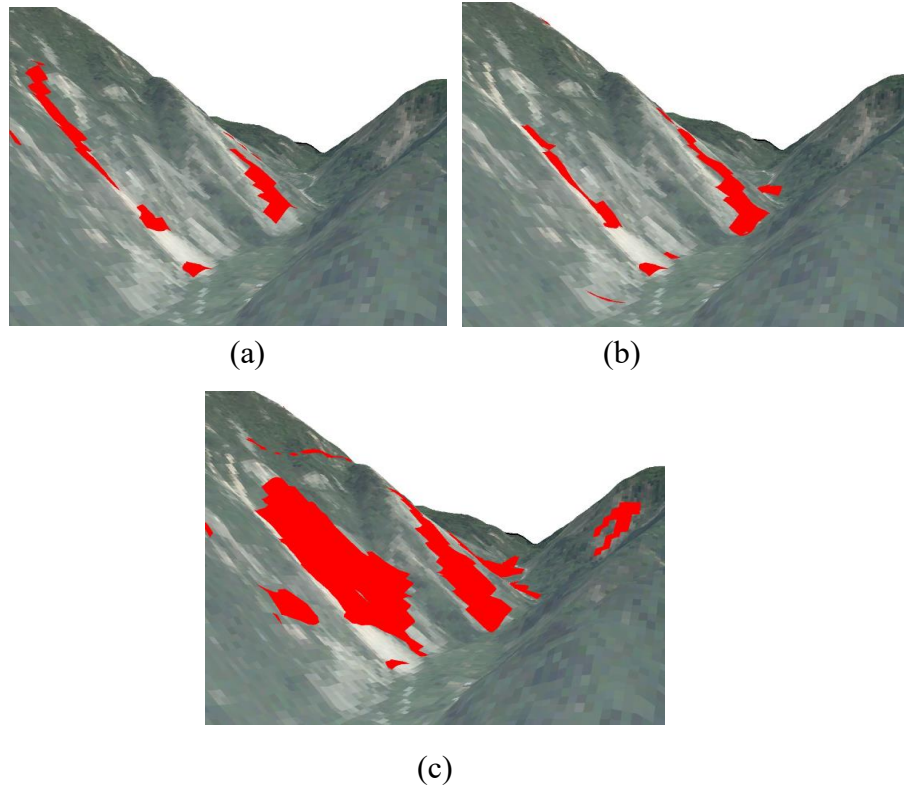

(c)

Figure 13. DECD results of sample area 2 obtained by (a) 3σR, (b) C3σR and (c) AC3σR.

*3.3 Discussion*

3σR is less accurate than AC3σR and C3σR because the influence of sample outliers (i.e., the changing elevation) is not considered when the outliers are removed via the 3σR, which results in samples contamination (including outliers) and inaccurately estimated parameters and leads to unsatisfactory change detection results. Therefore, based on the basic principle of moment estimation, AC3σR and C3σR were used to remove the outliers to obtain a higher accuracy than 3σR.

AC3σR is more robust than C3σR. Under different terrain steepness and registration error conditions, AC3σR has a change rule of the index value with an adaptive threshold; that is, the DECD accuracy of AC3σR in the convergence radius [0, 0.30010] remains stable and does not decline, and the censored ratio $β$ with the highest accuracy is close to the ratio of the changing grid number to the total grid number in a test area of 0.00399. However, the convergence radius of C3σR is only [0, 0.05010], and the convergence radius of AC3σR is much greater than the real change ratio of DEM and the convergence radius of C3σR. The reason for this is that the robust iterative estimation method can approach the optimal solution by performing multiple iterations so that the process of determining the censored ratio is less affected by the initial value yet still converges to the optimal accuracy in the radius of convergence, thus realizing DECD results with good stability.

AC3σR is more accurate than C3σR for the following reasons. First, robust iterative estimation has fault tolerance. According to (2), if the initial set of the censored samples proportion is not accurate, then the AC3σR method can correct or adjust it by iterative approximations; however, C3σR does not have this fault tolerance capability. Second, robust iterative estimation approaches the optimal solution through loop iterations; however, C3σR does not have loop processing. Therefore, C3σR is less accurate than AC3σR.

The global $F_{ZD}$ follows the approximate symmetric distribution after removing the censored samples and is generally not affected by local asymmetric distribution under conditions of high-accuracy DEM generation and high-accuracy DEM registration, which can be explained as follows: (1) with quality assurance, e.g., high-accuracy photogrammetric processing, DEM registration and censored samples, the accuracy of DEM generation is relatively high, which can ensure that the global $F_{ZD}$ follows approximately symmetric distribution; (2) for DEM change of the experimental area

Wenchuan county, the proportion of real DEM changes is very small relative to the invaried elevation. Within the allowable range of measurement error, the global $F_{ZD}$ follows the approximate symmetric distribution for invaried elevation, which is in line with the real DEM change of Wenchuan County between 2010 and 2011; (3) the larger the censored global sample size is, the more approximate the $F_{ZD}$ is to symmetric distribution. The global sample size is very large, and the local $F_{ZD}$ could be "averaged" by global $F_{ZD}$, so it is closer to the population. However, for local samples, the number of local $F_{ZD}$ is much smaller than that of global $F_{ZD}$ so that local samples may not represent the population, that is to say that the larger the scale is, the stronger the statistical law is. Moreover, the local areas of asymmetric distribution are usually detected as outliers and eliminated by the proposed AC3σR. In table 9, when the unilaterally censored samples (i.e., outliers) proportion is 11%, the absolute values of ∆ and $\hat{\mu}_3$ of $F_{ZD}$ are |-0.00219| and |0.00119| based on global AC3σR, which decreases from |$\hat{\mu}_{13}$= -0.41843|, | $\hat{\mu}_{23}$= 0.97693| and |∆$_1$= 0.29586|, |∆$_2$= 0.90837| based on local AC3σR in local area 1 and local area 2 respectively and from |$\hat{\mu}_{13}$= -0.71251|, | $\hat{\mu}_{23}$= 1.22597| and |∆$_1$= -2.60068|, |∆$_2$= 1.12836| without the proposed censoring in local area 1 and local area 2 respectively. Therefore, even if the local $F_{ZD}$ (i.e., local samples) does not follow the symmetric distribution, the approximate symmetric distribution of the global $F_{ZD}$ (i.e., global population) is not affected. Please note that global parameters ($\hat{\mu}$ and $\hat{\sigma}$) of the censored global samples are used to detect the local DEM changes.

The censored method (i.e., bilateral censoring or unilateral censoring) has a certain influence on the experimental results. For simulated experiment, due to the terrain data generated by Gaussian synthetic surface function, which has good spatial distribution and follows the symmetrical distribution i.e., Gaussian distribution, bilateral censoring for C3σR and AC3σR performs better than unilateral censoring. For real-world experiment, due to data disturbed by landslide and earthquake etc., which may lead to the inconsistency between the sample size of elevation increase and that of elevation decrease (i.e., non-conservation of energy in a basin) and follow the heavy tailed distribution, unilateral censoring for C3σR and AC3σR performs better than bilateral censoring. However, some outliers of DEM differencing samples censored and removed by unilateral censoring, the remaining samples are pollution-free samples so that it follows the approximate symmetrical distribution.

On the basis of AC3σR, the experimental results of DECD are basically consistent with the known areas. Based on the visual interpretation and the reported research results (Gan et al. 2012, Xu et al. 2012), the selected debris flow area in Figure 12 includes the accumulation area along the Minjiang River downstream of the Hongchun gully (N $31°04'01.1"$, E $103°29'32.7"$) (Han et al. 2013, Li et al. 2015, Chen et al. 2016) debris flow and two smaller debris flow accumulation areas. Figure 13 shows that the landslide risk areas and high-risk areas are mainly located on both sides of the river, especially in the southwest of the area, which is also consistent with the research results of other reports.

High-accuracy geographic registration is a prerequisite for DECD. In the simulation experiment, when the registration error increases, the registration error magnifies the result of DEM differencing as the elevation steepness increases, which results in many false changes. For example, a steep mountain area has a certain dislocation that leads to large DEM differences, and the DEM differencing samples may not satisfy the approximate symmetric distribution. Therefore, the DEM that has not changed is detected as a change area, resulting in a poor AC3σR effect. The reason why C3σR is slightly more accurate than AC3σR is that DEM differencing shows an irregular probability distribution at this time, which leads to the failure of the statistical method and reduces the experimental accuracy of DECD.

AC3σR has strong universality in real data experiments. In complex terrains, AC3σR, which is based on the principle of robust iterative moment estimation, can effectively remove the influence of outliers and reduce the initial value of the censored ratio on the influence of DECD. The accuracy of the real data experiment is better than that of the 3σR and C3σR methods, which verifies the high accuracy and robustness of AC3σR-based DECD.

## 4. Conclusions

This paper uses a degree-1 trivariate polynomial, the DEM differencing method and censored samples to quantitatively simulate the change detection uncertainty of a three-dimensional position. The experimental results show the following. (1) Positional uncertainty plays a large role in 3D change detection, and 3D registration errors may lead to pseudo-changes; therefore, high-accuracy geographic registration is a prerequisite for DECD. (2) The quality assurance strategy used for DECD based on AC3σR proposed in this paper has the highest accuracy and the strongest robustness in DECD; thus, (a) Under the condition of high-accuracy DEM generation and high-accuracy DEM registration, the proposed quality assurance strategy (i.e., AC3σR) can ensure that the global DEM differences follow the approximate symmetric distribution; (b) AC3σR can iteratively approximate and detect relatively real elevation changes within the radius of convergence and exhibits the highest accuracy; (3) compared with the traditional DECD method, AC3σR is less affected by the initial value of the censored samples proportion and has self-adaptability and, thus, very strong robustness; (4) unilateral and bilateral censoring for global AC3σR can also ensure approximate symmetric distribution and the accuracy of parameter estimation at different scenes.

Higher change-detection accuracy can be obtained through the proposed quality assurance strategy, i.e., AC3σR. A better understanding of this strategy would be significant for reducing the adverse effects of 3D misregistration and improving change detection accuracy. However, there are still two factors to be explored further. First, large 3D misregistration can cause the statistical distribution of DEM differences to disregard regularity. Thus, it is worth exploring whether a more effective means of quality assurance exists. Second, the quality assurance method of DECD is developed in this paper for geographic grids or pixels. Third, this paper studies a quality assurance strategy for positional uncertainty from the perspective of geometry but does not involve attribute change detection. In the future, research will focus on (1) the DECD of attributes with quality assurance by using methods that involve machine learning, such as the application of DECD in geological disasters (e.g., debris flows and landslides), and (2) the object-oriented method will be adopted to further overcome the interference of 3D misregistration in DECD and improve the accuracy of DECD.

**Acknowledgements**

The authors thank the National Natural Science Foundation of China (NSFC) (Grant No. 41771493 and 41101407) for supporting this work. The authors are grateful for the comments and contributions of the anonymous reviewers and the members of the editorial team.

**Data availability statement**

The simulation data and codes that support the findings of this study are available from Figshare at https://figshare.com/s/fff0b63893e35edf73ce.